\documentclass[conference]{IEEEtran}
\IEEEoverridecommandlockouts

\usepackage{amsmath, amssymb, amsfonts}
\usepackage[dvipsnames]{xcolor}
\usepackage{graphicx}
\usepackage{textcomp}

\usepackage{cite, flushend}
\usepackage{algorithmic}
\usepackage{comment}

\usepackage{tikz}
\usetikzlibrary{calc}
\usepackage{pgfplots, pgfplotstable}

\pgfplotsset{compat=1.18}

\usepackage{acronym}
\newacro{JSAC}{Joint sensing and communication}
  \newacro{UE}{User Equipment}
   \newacro{CPU}{central processing unit}
  \newacro{BF}{beamforming}
  \newacro{AP}{Access Point}
  \newacro{CF-MIMO}{Cell-Free Multiple-Input Multiple-Output}
   \newacro{SDR}{semidefinite relaxation}
 \newacro{CF-mMIMO}{Cell-Free massive MIMO}   
  \newacro{RIS}{Reconfigurable Intelligent Surfaces}
  \newacro{SCA}{Successive Convex Approximation}
  \newacro{ZF}{Zero Forcing}
  \newacro{SDP}{positive semidefinite}
  \newacro{MRT}{Maximum Ratio Transmission}

\usepackage{verbatim}
\usepackage{braket}
\usepackage{physics}
\usepackage{tabto}

\PassOptionsToPackage{symbol}{footmisc}
\usepackage{footmisc}
\usepackage{nomencl}

\renewcommand{\footnotesize}{\scriptsize}
\newtheorem{theo}{\textbf{Theorem}}
\newtheorem{lemma}{\textbf{Lemma}}

\newtheorem{definition}{\textbf{Definition}}

\def\BibTeX{{\rm B\kern-.05em{\sc i\kern-.025em b}\kern-.08em
    T\kern-.1667em\lower.7ex\hbox{E}\kern-.125emX}}

\begin{document}

\title{Beamforming Tradeoff for Sensing and Communication in Cell-Free MIMO\\
}

\author{\IEEEauthorblockN{Xi Ding and Luca Kunz and E. Jorswieck}
\IEEEauthorblockA{\textit{Institute for Communications Technology} \\
\textit{Technische Universität Braunschweig}\\
Braunschweig, Germany \\
\{xi.ding,luca.kunz,e.jorswieck\}@tu-braunschweig.de}
}

\maketitle

\begin{abstract}
This paper studies optimal joint \ac{BF} for \ac{JSAC} in small-scale \ac{CF-MIMO} systems. While prior works have explored JSAC optimization using methods such as \ac{SCA} and \ac{SDR}, many of these approaches either lack global optimality or require additional rank-reduction steps. In contrast, we propose an SDR-based optimization framework that guarantees globally optimal solutions without post-processing. To benchmark its performance, we introduce a standalone BF strategy that dedicates each \ac{AP} exclusively to either communication or sensing. The proposed formulation builds upon a general multi-user system model, enabling future extensions beyond the single-user setting. Overall, our framework offers a globally optimal and computationally efficient BF design, providing valuable insights for the development of next-generation wireless networks.
\end{abstract}


\section{Introduction}

As 5G communications mature, the focus is shifting to the design of next-generation 6G networks, driving the need to explore advanced technology paradigms.  
A key challenge in 6G network design is the efficient integration of JSAC with \ac{CF-mMIMO}\cite{bj} and \ac{RIS}\cite{overVRIS}, both of which are expected to play a transformative role in future wireless systems.  
JSAC enables the seamless convergence of communication and sensing \cite{caire, xiong}, using shared spectral and hardware resources to improve network efficiency while enabling real-time environmental sensing and data transmission \cite{overVzhang, demirhan}.

This integration is particularly crucial for applications such as vehicular networks \cite{yang, Nearfield_lin}, industrial automation, and smart cities \cite{smartC, overVkatwe}, where both high communication reliability and precise situational awareness are essential. Compared to separate communication and radar sensing systems, JSAC significantly improves spectral utilization and operational efficiency, making it a cornerstone of next-generation 6G networks \cite{70overVliu}.  
Compared to cellular massive MIMO, \ac{CF-mMIMO} improves coverage and mitigates inter-cell interference through coherent transmission from distributed \acp{AP}. However, its integration with \ac{JSAC} introduces significant beamforming challenges, particularly in mitigating interference among distributed \acp{AP}, necessitating advanced precoding techniques \cite{maOptBF}. 

\subsection{Existing Works}

Recent advancements in \ac{JSAC} have explored \ac{RIS}-enhanced communication, including active \ac{RIS} \cite{AktivRIS} and cell-free \ac{RIS}-assisted MIMO \cite{AidedRIS}. Studies on \ac{RIS}-based \ac{JSAC} systems have examined both passive and active \ac{RIS} designs, analyzing their impact on spectral efficiency and sensing resolution. Moreover, \ac{RIS} facilitates full-duplex  \ac{JSAC} by mitigating self-interference via optimized reflections, while simultaneously enhancing spectral efficiency and vehicle sensing. This enables the seamless integration of \ac{JSAC} into \ac{CF-MIMO} systems \cite{FDISAC}. 

In CF-mMIMO JSAC beamforming, \cite{mao} employs \ac{SCA} to obtain a locally optimal solution, providing a feasible solution but lacking global optimality. While \cite{de1}, the most relevant work to ours, explores different approaches, it shares the same \ac{JSAC} beamforming framework and utilizes SDR techniques to derive globally optimal closed-form beamforming structures \cite{luo}. 
The JSAC optimization primarily aims to maximize sensing $\mathrm{SNR}_{\mathrm{s}}$ while maintaining the communication $\mathrm{SINR}_{\mathrm{c}}$ above a fixed threshold, which limits broader trade-off exploration.

\subsection{Contributions} 

This paper investigates the weighted sum SNR optimization problem for \ac{JSAC} in small-scale \ac{CF-MIMO} systems. We propose an SDR-based optimization framework that guarantees a globally optimal rank-one solution without the need for additional rank-reduction techniques \cite{luo,huang}. The proposed framework applies the scalarization approach of the multi-objective programming \cite{emil14} and thereby simultaneously maximizes both sensing $\mathrm{SNR}_{\mathrm{s}}$ and communication $\mathrm{SNR}_{\mathrm{c}}$, ensuring an optimal trade-off between the two functionalities. The formulation builds upon a general multi-user system model that captures the JSAC trade-off and enables future extensions beyond the single-user case.
The proposed approach is evaluated against benchmark strategies and achieves the same performance as \ac{MRT} in both communication- and sensing-optimal regimes through joint design.
Additionally, a standalone beamforming strategy is introduced, where each \ac{AP} is assigned exclusively to either communication or sensing, illustrating the performance gains enabled by unified optimization. 
The \ac{ZF} strategy is also included as a baseline reference; while it remains feasible, it is not Pareto-optimal and holds limited relevance in the single-user setting. Theoretical analysis confirms the solution's feasibility and optimality, addressing key challenges in efficient \ac{JSAC} beamforming for \ac{CF-MIMO} systems.

\subsection{Organization and Notations}

The paper is structured as follows: Section~II presents the system model, outlining the \ac{JSAC} framework in \ac{CF-MIMO}, including both the communication and sensing models. Section~III formulates the beamforming optimization problem and introduces the \ac{SDR}-based solution. Section~IV presents numerical results comparing the proposed method with benchmark strategies, including \ac{MRT}, standalone beamforming, and \ac{ZF}. Sections~V and~VI conclude the paper with discussions on future directions. The mathematical foundations and proofs supporting the optimization framework are provided in the Appendix. 
In our formulation, all matrices are considered {bounded linear operators} on the Hilbert space \( \mathcal{H} \), as defined in {Appendix~\ref{LA}}. The norm \( \|\cdot\| \) refers to the {operator norm}, as defined in {Appendix~\ref{LA}}. In the case of scalars, the notation \( \left| \cdot \right| \) represents a simplified form of the operator norm.

\section{System Model}

We describe the \ac{JSAC} model in a \ac{CF-MIMO} system with a multi-user case (II-A) and a single-user setup (II-B).

\subsection{General multi-user model}\label{multiUE}

As shown in Fig.~\ref{fig_n}, we consider a multi-user \ac{CF-MIMO} system, where a central processing unit (CPU) connects to two distributed \acp{AP} via a high-frequency fronthaul link (e.g., mmWave). The CPU computes beamforming vectors to coordinate downlink transmission.
\begin{figure}[htbp]
    \centering
    \input{sysmod_n}
\end{figure}
Each \ac{AP} employs distinct beamformers to jointly serve \( K \) \acp{UE}, indexed by \( k \in \{1, 2, \dots, K\} \), and to simultaneously sense a vehicle, such as a vehicle, over a sub-6GHz channel.\footnote{
SINR\(_\mathrm{c}\) in \eqref{eq:SINR_comm} and SNR\(_\mathrm{s}\) in \eqref{eq:SNR_sensing} extend the single-user model in Section~II-B and appear in the objective in \eqref{ojf1}, from which \eqref{eq:joint_opt} and \eqref{eq:opt1} follow. For \(K = 1\), SINR\(_\mathrm{c}\) simplifies to SNR\(_\mathrm{c}\) since interference is absent.}

The communication $\mathrm{SNR}_{\mathrm{c}}$ at \ac{UE} \( k \) is given by
\begin{align}\label{eq:SINR_comm}
    \mathrm{SINR}_{\mathrm{c}}^{(k)} = 
    \frac{
        \left| \mathbf{h}_{1k}^\mathrm{H} \mathbf{w}_1^{(k)} + \mathbf{h}_{2k}^\mathrm{H} \mathbf{w}_{2}^{(k)} \right|^2
    }{
        \sigma_1^2 + 
        \sum\limits_{\ell \neq k} 
        \left| 
            \mathbf{h}_{1\ell}^\mathrm{H} \mathbf{w}_1^{(\ell)} 
            + 
            \mathbf{h}_{2\ell}^\mathrm{H} \mathbf{w}_2^{(\ell)} 
        \right|^2
    }.
\end{align}
The sensing $\mathrm{SNR}_{\mathrm{s}}$ at this vehicle is expressed as
\begin{align}\label{eq:SNR_sensing}
    \mathrm{SNR}_{\mathrm{s}} = 
    \|\mathbf{g}_2\|^2 \cdot
    \frac{
        \left| 
            \sum_{k=1}^{K} 
            \left(
                \mathbf{g}_1^\mathrm{H} \mathbf{w}_1^{(k)} 
                + 
                \mathbf{g}_2^\mathrm{H} \mathbf{w}_2^{(k)} 
            \right) 
        \right|^2
    }{
        \sigma_2^2
    }.
\end{align}
To balance the communication and sensing performance, we first define the joint optimization problem as:
\begin{subequations}\label{eq:joint_opt}
\begin{align}
&\max_{\mathbf{w}_1, \mathbf{w}_2} \quad 
 \alpha \sum_{k=1}^{K} \mathrm{SINR}_\mathrm{c}^{(k)} 
+ \overline{\alpha} \cdot \mathrm{SNR}_{\mathrm{s}} \\
&\text{s.t.} \quad 
 \left\|\mathbf{w}_1\right\| = \left\|\mathbf{w}_2\right\| \leq 1.
\end{align}
\end{subequations}
Building on the trade-off structure in~\eqref{eq:joint_opt}, we derive the corresponding multi-user formulation:
\begin{subequations}\label{eq:opt1}
\begin{align}
&\max_{\mathbf{w}_1,\mathbf{w}_2} \quad 
 \alpha\sum_{k=1}^{K} \frac{
        \left| \mathbf{h}_{1k}^\mathrm{H} \mathbf{w}_1^{(k)} + \mathbf{h}_{2k}^\mathrm{H} \mathbf{w}_{2}^{(k)} \right|^2
    }{
        \sigma_1^2 
        + 
            \sum\limits_{\ell \neq k} 
            \left| 
                \mathbf{h}_{1\ell}^\mathrm{H} \mathbf{w}_1^{(\ell)} 
                + 
                \mathbf{h}_{2\ell}^\mathrm{H} \mathbf{w}_2^{(\ell)} 
            \right|^2
            } \notag \\
\quad &\quad + \overline{\alpha}  \|\mathbf{g}_2\|^2 
    \frac{
        \left| 
            \sum_{k=1}^{K} 
            \left(
                \mathbf{g}_1^\mathrm{H} \mathbf{w}_1^{(k)} 
                + 
                \mathbf{g}_2^\mathrm{H} \mathbf{w}_2^{(k)} 
            \right) 
        \right|^2
    }{
        \sigma_2^2  } \\
&\text{s.t.} \quad 
 \left\|\mathbf{w}_1 \right\| = \left\|\mathbf{w}_2 \right\| \leq 1.
\end{align}
\end{subequations}
This objective captures aggregate communication performance across all \acp{UE} while preserving the sensing–communication trade-off. Its structure aligns with the formulation in~\cite{de1}, and a detailed comparison based on our solution is provided in a separate manuscript under preparation.

We consider the single-user case as a simplified setup for tractable optimization, as illustrated in Fig.~\ref{fig1}.

\subsection{Simplified Single-User Case}\label{VereinfachendesM}

\begin{figure}[htbp]
    \centering
    \input{sysmod}
\end{figure}
Based on the setup illustrated in Fig.~\ref{fig1}, we next detail the signal models for communication and sensing.

\subsubsection{Communication Model}
In the communication system, the signal  ${y}_1\in \mathbb{C}$ received at the UE is expressed as:
\begin{align}
    {y}_1 = \mathbf{h}_{1}^\mathrm{H}  \mathbf{w}_1 {s}_1 + \mathbf{h}_2^\mathrm{H}  \mathbf{w}_2 {s}_2 + {n}_1
\end{align}
 where \( {s}_1, {s}_2 \in \mathbb{C} \) represent the transmitted signals intended for the \ac{UE}, while \( \mathbf{h}_{1},\mathbf{h}_{2} \in  \mathcal{H}\) denote the corresponding channel vectors between AP 1 and AP 2 and the \ac{UE}, respectively. \( \mathbf{w}_{1}, \mathbf{w}_{2} \in  \mathcal{H}\) are the beamforming vectors computed at the CPU and applied at the \acp{AP} to direct signals toward the UE. Finally, \( n_1 \) denotes the receiver noise at the \ac{UE}, modeled as a complex Gaussian random variable with zero mean and variance \( \sigma_1^2 \), i.e., \( n_1 \sim \mathcal{CN}(0, \sigma_1^2) \) with $\sigma_1 \in \mathbb{R}$. 

Each \ac{AP} is subject to a power constraint that limits its expected transmit power to a predefined maximum:
\begin{equation}
    \mathbb{E}\left[ \| \mathbf{w}_{1} {s}_1 \|^2 \right] \leq {P}_{1,\max}, \quad
    \mathbb{E}\left[ \| \mathbf{w}_{2} {s}_2 \|^2 \right] \leq {P}_{2,\max}.
\end{equation}
We assume finite transmit power limits \( P_{1,\max}, P_{2,\max} < \infty \) and normalized transmit signals, i.e.,
\begin{align}
    \mathbb{E}\left[ |s_1|^2 \right] = \mathbb{E}\left[ |s_2|^2 \right] = 1. \label{s_cons}
\end{align}
Under this assumption, the power constraints on the beamformers \( \mathbf{w}_{1}  \) and \( \mathbf{w}_{2}  \) simplify to:  
\begin{align}
    \| \mathbf{w}_{1} \|^2 \leq \mathrm{P}_{1,\max}, \quad
    \| \mathbf{w}_{2}  \|^2 \leq \mathrm{P}_{2,\max}. \label{w12cons}
\end{align}
Under ideal network synchronization, in a distributed MIMO system, a joint transmission strategy is employed, where multiple \acp{AP} cooperatively transmit the same data stream, i.e., \( {s}_1 = {s}_2 = {s} \). Each \ac{AP} applies a different beamforming vector, \( \mathbf{w}_{1}  \) and \( \mathbf{w}_{2}  \), allowing the signals to be adjusted through beamforming so that they are constructively combined at the UE, resulting in the following $\mathrm{SNR}_{\mathrm{c}}$ expression:
\begin{align}
    \mathrm{SNR}_{\mathrm{c}}(\mathbf{w}_1,\mathbf{w}_2) &=\frac{\mathbb{E}\left[\left|(\mathbf{h}_1^\mathrm{H}  \mathbf{w}_1 + \mathbf{h}_2^\mathrm{H}  \mathbf{w}_2){s}\right|^2\right]}{\mathbb{E}\left[ \left |\mathrm{n}_1 \right |^2  \right]}\notag\\
    &\overset{\eqref{s_cons}}{=}\frac{\left|(\mathbf{h}_1^\mathrm{H}  \mathbf{w}_1 + \mathbf{h}_2^\mathrm{H}  \mathbf{w}_2) \right|^2}{\sigma_1^2}. \label{SINR_c}
\end{align}

 \subsubsection{Sensing Model}
 
In the sensing system, the signal ${y}_2\in\mathbb{C} $ received at the vehicle is given by:
\begin{align}
    {y}_2 = (\mathbf{g}_{1}^\mathrm{H}  \mathbf{w}_1 {s}) \mathbf{g}_2^\mathrm{H}  \mathbf{v}_2 + (\mathbf{g}_2^\mathrm{H}  \mathbf{w}_2 {s}) \mathbf{g}_2^\mathrm{H}  \mathbf{v}_2 + {n}_2,
\end{align}
where
$s$ is the common transmitted symbol under the Joint Transmission strategy and \( \mathbf{g}_1,\mathbf{g}_2 \in  \mathcal{H}\) denote the channel vectors between the APs and the vehicle. Specifically, \( \mathbf{g}_1 \) represents the channel between \ac{AP}~1 and the vehicle, while \( \mathbf{g}_2 \) corresponds to the channel between \ac{AP}~2 and the vehicle.  
The term \( \mathbf{g}_2^\mathrm{H}  \mathbf{v}_2 \in \mathbb{C} \) represents the sensing beamforming operation at \ac{AP}~2, which enhances the reception of the reflected sensing signal. Here, \( \mathbf{v}_2 \in  \mathcal{H}\) is the beamforming vector applied at \ac{AP}~2 to optimize the received sensing signal.  
Finally, \( n_2 \) denotes the sensing noise, modeled as \( \mathcal{CN}(0, \sigma_2^2) \) with $\sigma_2 \in \mathbb{R}$.

To maximize the received reflected signal at \ac{AP}~2, 
\( \mathbf{v}_2 \) is aligned with the received signal direction, yielding:
\begin{align}
    \mathbf{v}_2 = \frac{\mathbf{g}_2 \overline{(\mathbf{g}_1^\mathrm{H}  \mathbf{w}_1 + \mathbf{g}_2^\mathrm{H}  \mathbf{w}_2)}}{ \left\| \mathbf{g}_2 \overline{(\mathbf{g}_1^\mathrm{H}  \mathbf{w}_1 + \mathbf{g}_2^\mathrm{H}  \mathbf{w}_2)}\right\|}.\label{v2}
\end{align}
The sensing \( \mathrm{SNR}_{\mathrm{s}} \) as a function of \( \mathbf{w}_1 \) and \( \mathbf{w}_2 \) is given by:
\begin{align}
    \mathrm{SNR}_{\mathrm{s}}(\mathbf{w}_1,\mathbf{w}_2)&=\frac{\mathbb{E}\left[  \left\|(\mathbf{g}_1^\mathrm{H}  \mathbf{w}_1 s + \mathbf{g}_2^\mathrm{H}  \mathbf{w}_2 s) \mathbf{g}_2^\mathrm{H}  \mathbf{v}_2 \right\|^2\right]}{\mathbb{E}[\left |{n}_2\right |^2]}\notag\\
   &\overset{\eqref{v2}}{=}\frac{\left\|\mathbf{g}_2 \right\|^2 \left| \mathbf{g}_1^\mathrm{H} \mathbf{w}_1 + \mathbf{g}_2^\mathrm{H}  \mathbf{w}_2 \right|^2}{ \sigma_2 ^2}. \label{SNR_s}
\end{align}

\section{Problem Statement and Main Result}
 We first define the following objective function:  
\begin{equation}
    {f}(\mathbf{w}_1, \mathbf{w}_2) = \alpha\ \mathrm{SNR}_\mathrm{c} + \overline{\alpha}\ \mathrm{SNR}_\mathrm{s}, \label{ojf1}
\end{equation}  
where $\alpha \in [0,1] $ and $ \overline{\alpha}=1-\alpha$ are the predefined weighting factors that balances the trade-off between communication performance, characterized by \( \mathrm{SNR}_{\mathrm{c}} \) in \eqref{SINR_c}, and sensing performance, represented by \( \mathrm{SNR}_{\mathrm{s}} \) in \eqref{SNR_s}. 

    Since we are primarily concerned with the trade-off between communication performance and sensing performance, the noise power in \eqref{SINR_c} and \eqref{SNR_s} is not a critical factor in our analysis. Without loss of generality, we set the noise power to 1 to simplify the formulation in \eqref{SINR_c} and \eqref{SNR_s}, resulting in the objective function $f(\mathbf{w}_1, \mathbf{w}_2)$ as follows: 
\begin{align}
    f(\mathbf{w}_1,\mathbf{w}_2)&= \alpha  \left | \mathbf{w}_1^\mathrm{H} \mathbf{h}_1 + \mathbf{w}_2^\mathrm{H}  \mathbf{h}_2 \right |^2 \notag\\
    &+ \overline{\alpha}  \left\| \mathbf{g}_2\right\|^2 \left | \mathbf{w}_1^\mathrm{H}  \mathbf{g}_1 + \mathbf{w}_2^\mathrm{H} \mathbf{g}_2 \right |^2 \notag\\
    &= \alpha  \left | \mathbf{w}_1^\mathrm{H}  \mathbf{h}_1 + \mathbf{w}_2^\mathrm{H}  \mathbf{h}_2 \right |^2  +\widetilde{\alpha}   \left | \mathbf{w}_1^\mathrm{H}  \mathbf{g}_1 + \mathbf{w}_2^\mathrm{H}  \mathbf{g}_2 \right |^2, \label{OP_n}
\end{align}
 with $\widetilde{\alpha}=\overline{\alpha}  \left\| \mathbf{g}_2\right\|^2$. The extension to arbitrary noise powers \( \sigma_1^2 \) and \( \sigma_2^2 \) is straightforward. Our goal is to maximize this objective function in \eqref{OP_n} under the constraints in \eqref{w12cons} with ${P}_{1,\max}={P}_{2,\max}=1$:
\begin{align}
   \left\|\mathbf{w}_1 \right\|^2 = \left\|\mathbf{w}_2\right\|^2 \leq  1.
\end{align}  
Formally, the optimization problem is expressed as:
\begin{subequations}\label{OP_newFormulate}
    \begin{align}
    &\max_{\mathbf{w}_1, \mathbf{w}_2} {\alpha  \left | \mathbf{w}_1^\mathrm{H}  \mathbf{h}_1 + \mathbf{w}_2^\mathrm{H}  \mathbf{h}_2  \right |^2 +
\widetilde{\alpha}\left |\mathbf{w}_1^\mathrm{H}  \mathbf{g}_1 + \mathbf{w}_2^\mathrm{H}  \mathbf{g}_2 \right |^2 },\label{OPT_o}\\
   & \quad \mathrm{s.t.} \quad \left\|\mathbf{w}_1 \right\| = \left\|\mathbf{w}_2 \right\| \leq 1. 
\end{align} 
\end{subequations}
In \eqref{OPT_o}, it is not immediately clear whether this optimization problem is jointly convex in $(\mathbf{w}_1, \mathbf{w}_2)$. While we can determine convexity by computing the Hessian matrix, as shown in \cite{maOptBF}, this approach, although straightforward in principle, becomes impractical for complex objective functions due to the extensive analysis required. Therefore, to simplify the process, we transform our optimization variables using the following approach.
To clearly distinguish the transformation in the optimization problem before and after variable substitution, we highlight the updated objective function and variables in bold. 
We can further simplify the objective function in \eqref{OPT_o} by expressing the beamforming vectors \( \mathbf{w}_1 \) and \( \mathbf{w}_2 \) as a single vector, denoted as \( \mathbf{w} \), where  
\begin{align}
    \mathbf{w} = (\mathbf{w}_1,\mathbf{w}_2) \in  \mathcal{H}.
\end{align}  
Similarly, the channel coefficients \( \mathbf{h}_1 \) and \( \mathbf{h}_2 \) can be combined into a unified vector, denoted as \( \mathbf{h} \), such that  
\begin{align}
    \mathbf{h} = (\mathbf{h}_1,\mathbf{h}_2) \in \mathcal{H}.
\end{align}  
Likewise, the sensing-related channel coefficients \( \mathbf{g}_1 \) and \( \mathbf{g}_2 \) can be grouped into another vector, denoted as \( \mathbf{g} \), given by  
\begin{align}
    \mathbf{g}  = (\mathbf{g}_1,\mathbf{g}_2) \in \mathcal{H}.    
\end{align} 
Using these vectors, the objective function is reformulated as:
\begin{subequations}\label{hgm}
    \begin{align}
    {{f}(\mathbf{w})} &= \alpha \left|  \mathbf {w}^\mathrm{H}  \mathbf {h} \right|^2 +  \widetilde{\alpha}\left| \mathbf {w}^\mathrm{H}  \mathbf {g}\right|^2 \in \mathbb{R} \label{obj_n}\\
    &=\Tr{ \alpha \left|  \mathbf {w}^\mathrm{H}  \mathbf {h} \right|^2 +  \widetilde{\alpha}\left| \mathbf {w}^\mathrm{H}  \mathbf {g}\right|^2}\in \mathbb{R} \label{obj_Tr1}\\
    &=\Tr{\alpha \mathbf {w}^\mathrm{H}  \mathbf {h}\mathbf {h}^\mathrm{H} \mathbf {w}}+\Tr{\widetilde{\alpha}\mathbf {w}^\mathrm{H}  \mathbf {g}\mathbf {g}^\mathrm{H} \mathbf {w}}\notag\\
    &=\alpha \Tr{\mathbf {w}^\mathrm{H}  \mathbf {H}\mathbf {w} }+\widetilde{\alpha}\Tr{\mathbf {w}^\mathrm{H}  \mathbf {G}\mathbf {w}}\label{obj_Tr2}\\
    &=\alpha\Tr{\mathbf {H} \mathbf {W}}+\widetilde{\alpha}\Tr{\mathbf {G} \mathbf {W}}\label{obj_Tr3}\\
    &=\Tr{(\alpha\mathbf {H}+\widetilde{\alpha}\mathbf {G})\mathbf {W} }\notag\\
    &=\Tr{\mathbf {M}\mathbf {W} }, \label{obj_end}
\end{align}  
\end{subequations}
where all the operators $\mathbf {H}=\mathbf {h}\mathbf {h}^\mathrm{H} $, $  \mathbf {W}=\mathbf {w}\mathbf {w}^\mathrm{H} $, $\mathbf {G}=\mathbf {g}\mathbf {g}^\mathrm{H} $, $ \mathbf {M}=\alpha\mathbf {H}+\widetilde{\alpha}\mathbf {G}\in \mathcal{B}(\mathcal{H})$ are bounded and linear, as defined in Definition~\ref{Def_BOP} in Appendix~\ref{LA}, and Hermitian, as stated in Def.~\ref{Def_HOP} in Appendix~\ref{LA}. Moreover, $\mathbf {H},\mathbf {W}$ and $\mathbf {G}$ are of rank one.
The transition from (\ref{obj_n}) to (\ref{obj_Tr1}) is justified by the fact that our objective function produces a real number.
Since the trace operation applied to a scalar returns the scalar itself in \eqref{TrDef}, we rewrite it in trace form for consistency.
Furthermore, the transformation from (\ref{obj_Tr2}) to (\ref{obj_Tr3}) follows from a fundamental property of the trace operation, as stated in Theorem~\ref{trace_properties} in Appendix~\ref{LA}. 
Due to the linearity of the trace\footnote{As stated in \cite[Satz 3.5.2]{huppert2010lineare}, the trace is a \( \mathbb{C}- \)linear mapping on the space of \( n \times n \) operators.}it follows directly that the final transformation in (\ref{obj_end}) holds.
Thus, we reformulate the optimization problem in \eqref{OP_newFormulate} based on \eqref{hgm} for \( {f}(\mathbf{w}) \) as follows:

\begin{subequations}\label{OPT_Rank1}
    \begin{align}
        &\max_{\mathbf{W}} \quad f(\mathbf{w}) 
        \overset{\eqref{obj_end}}{=} \max_{\mathbf{W}} \quad \Tr(\mathbf{M}\mathbf{W}), \\
        &\text{s.t.} \quad \Tr\left( \mathbf{W}
        \begin{bmatrix}
            \mathbf{I}_N & 0 \\
            0 & 0
        \end{bmatrix} \right) \leq 1, \\
        &\phantom{\text{s.t.} \quad} \Tr\left( \mathbf{W}
        \begin{bmatrix}
            0 & 0 \\
            0 & \mathbf{I}_N
        \end{bmatrix} \right) \leq 1, \\
        &\phantom{\text{s.t.} \quad} \mathbf{W} \succeq \mathbf{0}, \\
        &\phantom{\text{s.t.} \quad} \operatorname{rank}(\mathbf{W}) = 1,\label{delRank1}
    \end{align}
\end{subequations}
where \( \mathbf{I}_N \) denotes the \( N \times N \) identity operator. Being rank-one and Hermitian, $\mathbf{W}$ is \ac{SDP} with a single nonzero eigenvalue.
However, the rank-one constraint in~\eqref{OPT_Rank1} makes the problem non-convex. To handle this, we apply \ac{SDR} for handling non-convex quadratic problems~\cite{huang,luo}.

Since the rank function is non-convex, directly solving the problem with this constraint is intractable. 
To address this, we relax the rank-one constraint and allow $\mathbf{W} \succeq 0$.
This reformulates the problem into a standard SDP\footnote{The objective function in \eqref{hgm} is linear in $\mathbf{W}$.} problem, which is convex and can be efficiently solved using existing optimization tools. By eliminating the rank-one constraint in~\eqref{delRank1}, the problem in~\eqref{OPT_Rank1} is relaxed into a convex SDP, which can be solved efficiently.

Removing the rank-one constraint expands the feasible solution space. However, semidefinite optimization theory suggests that this relaxation has little effect on optimality. In practice, the solution to the relaxed problem often yields a near-optimal or even exact rank-one solution~\cite{luo}.
By applying problem ($\mathrm{P\,1}$) in \eqref{P1} (Appendix~\ref{Appen_SRO}) with \( L = 1 \) and \( M = 2 \), and substituting into the inequality \eqref{rankinq} in Lemma~\ref{lamm3.1}, we obtain the explicit bound:
\begin{align}
     \operatorname{rank}(\mathbf{W}^*) < \sqrt{2}.
\end{align}

Since the rank must be a non-negative integer and the trivial case $\operatorname{rank}(\mathbf{W}^*) = 0$ is excluded, it follows that \text{$\operatorname{rank}(\mathbf{W}^*) = 1$}. 
Hence, the optimal beamformer $\mathbf{W}^*$ can be determined directly and efficiently.

\section{Numerical Simulation and Assessment}

In this section, we present numerical simulations to validate the theoretical results established in previous sections. The optimization problem is implemented using Python, employing \ac{SDR} techniques to determine the optimal beamforming operator $\mathbf{W}^*$. The fixed channels $\mathbf{h}_1, \mathbf{h}_2, \mathbf{g}_1, \mathbf{g}_2 \in \mathbb{C}^3$ are randomly generated as Gaussian-distributed vectors and remain constant throughout the simulations.

\begin{figure}[htbp]
    \centering
\begin{tikzpicture}[scale=0.8]
    \begin{axis}
    [
        xlabel={Communication $\mathrm{SNR}_{\mathrm{c}}$ [dB]},
        ylabel={Sensing $\mathrm{SNR}_{\mathrm{s}}$ [dB]},
        title={Achievable $(\mathrm{SNR}_{\mathrm{s}}, \mathrm{SNR}_{\mathrm{c}})$ Region for Sensing and Communication},
        xmin=0,
        ymin=0,
        grid=major,
        legend style={
            at={(0.3,0.3)},
            anchor=south,
            font=\scriptsize,
            cells={anchor=west},
            fill=white,
            fill opacity=0.85,
            draw=black!40,
            line width=0.3pt,
            rounded corners=5pt, 
            inner xsep=0pt,
            inner ysep=0pt,
            row sep=0pt
        }
    ]
    \node[anchor=south west] at (axis cs: 1.5, 11) { $\alpha = 0$};
\draw[->,line width=0.25mm] (axis cs: 2, 12) -- (axis cs: 1, 13.3);  

\node[anchor=north east] at (axis cs: 6, 1.8) { $\alpha = 1$};
\draw[->,line width=0.25mm] (axis cs: 6, 1) -- (axis cs: 6.8, 2);  

\addplot[color=blue, line width=2pt, forget plot]
    table [col sep=comma] {SNRregion.csv};
\addlegendimage{legend image code/.code={
  \draw[blue, line width=2pt] (0.2cm,0cm) -- (0.4cm,0cm);
}}
\addlegendentry{Achievable ($\mathrm{SNR}_{\mathrm{s}}, \mathrm{SNR}_{\mathrm{c}}$) Region}

        \addplot[color=Orange, only marks, mark=diamond*, mark size=3pt] table [col sep=comma] {SNR_points.csv};
        \addlegendentry{Optimal Trade-off Points} 
        
        \addplot[color=Red, only marks, mark=*, mark size=3pt] table [col sep=comma] {SNRa.csv};
        \addlegendentry{MRT for Communication} 
        
        \addplot[color=Cyan, only marks, mark=*, mark size=3pt] table [col sep=comma] {SNRb.csv};
        \addlegendentry{MRT for Sensing} 
        
        \addplot[color=Purple, only marks, mark=square*, mark size=3pt] table [col sep=comma] {SNR_ZF.csv};
        \addlegendentry{Zero-Forcing BF} 
        
        \addplot[color=Green, only marks, mark=triangle*, mark size=3pt] table [col sep=comma] {SNR_base.csv};
        \addlegendentry{Standalone BF} 
    \end{axis}
\end{tikzpicture}
\caption{Performance comparison of different beamforming techniques for fixed channels. The channels are randomly generated as Gaussian-distributed vectors with dimension 3.}
\label{Ploten1}
\end{figure}
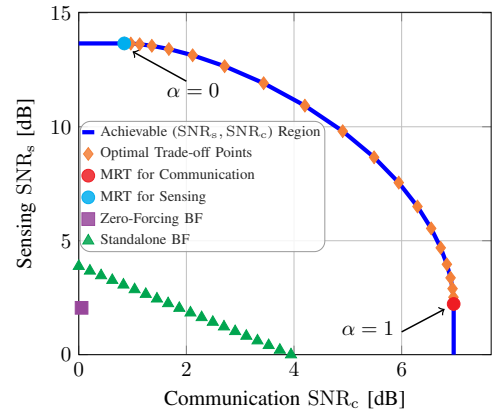

The results are visualized in Fig.~\ref{Ploten1}, where:
\begin{itemize}
 \item \textbf{Orange rhombuses}  represent the optimal trade-off points obtained via \ac{SDR}-based beamforming. 
 
\item The \textbf{blue curve}, formed by connecting these points, represents the achievable $(\mathrm{SNR}_\mathrm{s}, \mathrm{SNR}_\mathrm{c})$ region.

    \item \textbf{Green triangles} denote the Standalone BF strategy, where \ac{AP}~1 serves the \ac{UE} and \ac{AP}~2 performs sensing.

    \item \textbf{A red point} denotes the \ac{MRT} solution for communication, where beamforming maximizes $\mathrm{SNR}_{\mathrm{c}}$ at the \ac{UE}.
    
    \item \textbf{A cyan point} denotes the \ac{MRT} solution for sensing, where beamforming maximizes $\mathrm{SNR}_{\mathrm{s}}$ at the vehicle.

    \item \textbf{A purple square} marks the \ac{ZF} solution, where interference between \acp{AP} is completely eliminated.
\end{itemize}
\paragraph{Standalone Beamforming strategy}
The Standalone Beamforming strategy (green triangles) serves as a baseline reference for evaluating more sophisticated beamforming schemes. It represents a scenario where \ac{AP}~1 employs a direct beamforming vector aligned with its channel to the \ac{UE}, while \ac{AP}~1 applies a separate beamforming vector targeting only the sensing objective. Mathematically, this is formulated as:
\begin{equation}
    \mathbf{w}_1 = \frac{\mathbf{h}_1}{\| \mathbf{h}_1 \|}, \quad \mathbf{w}_2 = \frac{\mathbf{g}_2}{\|  \mathbf{g}_2\|}.
\end{equation}
Under this setup, the resulting $\mathrm{SNR}$ for communication and sensing are given by:
\begin{align}
    \mathrm{SNR}_{\mathrm{c}} = \| \mathbf{h}_1 \|^2, \quad
    \mathrm{SNR}_{\mathrm{s}} = \| \mathbf{g}_2 \|^4.
\end{align}
This setup simplifies the beamforming structure and highlights the performance potential of \ac{JSAC} optimization. 
Simulation results show that \ac{SDR}-based beamforming (blue region) outperforms standalone approaches by better balancing sensing and communication. 
Although suboptimal, standalone beamforming serves as an intuitive lower bound, revealing the trade-off without joint \ac{JSAC} optimization.
\paragraph{\ac{MRT} Beaforming}
\ac{MRT} focuses on signal power maximization at a single user, ignoring inter-user interference.
\begin{enumerate}
    \item For \( \alpha = 1 \), the optimal beamforming vectors are given by
    $\mathbf{w}_1 = \frac{\mathbf{h}_1 \sqrt{{P}_{1,\max}}}{\|\mathbf{h}_1\|}, \quad
    \mathbf{w}_2 = \frac{\mathbf{h}_2 \sqrt{{P}_{2,\max}}}{\|\mathbf{h}_2\|}$.
    \item For \( \alpha = 0 \), the optimal beamforming vectors are given by
    $\mathbf{w}_1 = \frac{\mathbf{g}_1 \sqrt{{P}_{1,\max}}}{\|\mathbf{g}_1\|}, \quad
    \mathbf{w}_2 = \frac{\mathbf{g}_2 \sqrt{{P}_{2,\max}}}{\|\mathbf{g}_2\|}$.
\end{enumerate}
We set \( P_{1,\max} = P_{2,\max} = 1 \) for normalization.

As expected, the two \ac{MRT}-based points align with the maximal achievable values of \( \mathrm{SNR}_{\mathrm{c}} \) and \( \mathrm{SNR}_{\mathrm{s}} \), corresponding to the communication-centric and sensing-centric cases, respectively. When beamforming is solely focused on sensing (\( \alpha = 0 \)) in Fig.~\ref{Ploten1}, the sensing $\mathrm{SNR}_{\mathrm{s}}$ reaches its peak. Conversely, when the focus is on communication (\( \alpha = 1 \)) in Fig.~\ref{Ploten1}, the communication $\mathrm{SNR}_{\mathrm{c}}$ attains its maximum.

\paragraph{\ac{ZF} Beamforming}
As our model involves only a single \ac{UE}, \ac{ZF} is included purely as a reference baseline.
To compute the \ac{ZF} beamforming vectors, we apply the pseudo-inverse of the channel matrix \( \mathbf{H} \), yielding the baseline beamformers $\mathbf{w_1^{\mathrm{ZF}}}$ and $\mathbf{w_2^{\mathrm{ZF}}}$. 
These vectors satisfy the orthogonality conditions
\[
\mathbf{w_1^{\mathrm{ZF}}} \perp \mathbf{h}_2, \quad \mathbf{w_2^{\mathrm{ZF}}} \perp \mathbf{h}_1,
\]
which eliminate inter-user interference. 
By substituting \( \mathbf{w_1^{ZF}} \) and \( \mathbf{w_2^{ZF}}\) into \eqref{SINR_c} and \eqref{SNR_s}, we verify that the \ac{ZF} solution falls within the achievable region in Fig.~\ref{Ploten1}. Simulation results indicate that \ac{SDR}-based beamforming performs comparably to \ac{MRT} across the trade-off range, while outperforming \ac{ZF} in interference-free scenarios, highlighting the robustness of the semidefinite relaxation approach.

\section{Conclusion}

This paper explores the trade-off between sensing and communication in small-scale \ac{CF-MIMO} systems by formulating an optimal beamforming problem.  We begin by introducing a general multi-user system model that characterizes the \ac{JSAC} trade-off, which also serves as the foundation for future multi-user optimization. We leverage an \ac{SDR}-based convex optimization framework to obtain a globally optimal beamforming solution. Our theoretical analysis ensures that the optimal solution is always rank-one, eliminating the need for additional rank-reduction techniques.
Numerical simulations validate the effectiveness of our approach by comparing \ac{SDR}-based beamforming with benchmark strategies. 
The results show that our method achieves \ac{MRT}-level performance in both communication- and sensing-optimal cases, while enabling smooth trade-offs between them through joint design. 
We further consider a standalone beamforming strategy, where each \ac{AP} is dedicated exclusively to either communication or sensing. This provides a reference to quantify the performance gain achieved by jointly optimizing both objectives. 
Additionally, the \ac{ZF} solution is included solely as a baseline, as it remains within the achievable region but is not Pareto-optimal.
Overall, this study presents an optimal and computationally efficient beamforming strategy for \ac{JSAC}, providing a robust and scalable solution for next-generation wireless networks.

\section*{Appendix}\label{Appendix}
\subsection{Background in Linear Algebra}\label{LA}
To rigorously formulate our optimization problem, we first establish the mathematical framework in which all matrices operate.  
We consider the finite-dimensional Hilbert space \( \mathcal{H} = \mathbb{C}^{N} \) over the field \( \mathbb{C} \), as defined in \cite[Definition 8.1.3]{huppert2010lineare}, equipped with the scalar product \( \langle \cdot, \cdot \rangle \), which induces the norm  
\begin{align*}
    \| \mathbf{x} \| = \sqrt{\langle \mathbf{x}, \mathbf{x} \rangle}, \quad \forall \mathbf{x} \in \mathcal{H}.
\end{align*}
Since the norm is induced by a scalar product, our Hilbert space is a normed space.
\begin{definition}[Bounded Linear Operator {\cite[pp. 142]{alt}}]\label{Def_BOP}
Let \( \mathcal{H} \) be a Hilbert space, and let \( \mathbf{A}: \mathcal{H} \to \mathcal{H} \) be a mapping. We define the space of bounded linear operators as  
\begin{align}
    \mathcal{B}(\mathcal{H}) := \{ \mathbf{A} \mid \mathbf{A} \textnormal{ is linear and continuous} \}.
\end{align}
All elements of $ \mathcal{B}(\mathcal{H})\cong \mathbb{C}^{{N}\times {N}}$ are called bounded linear~operators.
\end{definition}
\begin{definition}[Operator Norm {\cite[Satz II.1.4]{werner}}]\label{Def_OPN}
For any bounded linear operator \( \mathbf{A} \in \mathcal{B}(\mathcal{H}) \), the {operator norm} is defined as:
\begin{align}
    \|\mathbf{A}\| = \sup_{\|\mathbf{x}\| \leq 1} \|\mathbf{A} \mathbf{x}\| < \infty. \notag
\end{align}
\end{definition}    
\begin{definition}[Hermitian Operator {\cite[Satz V.5.5 ]{werner}}]\label{Def_HOP}
A bounded linear operator \( \mathbf{A} \in \mathcal{B}(\mathcal{H}) \) is called {Hermitian} if it satisfies:
\begin{equation}
    \langle \mathbf{A} \mathbf{x}, \mathbf{y} \rangle = \langle \mathbf{x}, \mathbf{A} \mathbf{y} \rangle, \quad \forall \mathbf{x}, \mathbf{y} \in \mathcal{H}.
\end{equation}
\end{definition}
In mathematics, the scalar product \( \langle \mathbf{A} \mathbf{x}, \mathbf{y}  \rangle \) corresponds to the engineering notation \( \mathbf{x}^\mathrm{H} \mathbf{A}^\mathrm{H}\mathbf{y}\), i.e.,  
\begin{align}
    \langle \mathbf{A} \mathbf{x}, \mathbf{y}  \rangle \triangleq (\mathbf{A}\mathbf{x})^\mathrm{H} \mathbf{y} = \mathbf{x}^\mathrm{H}\mathbf{A}^\mathrm{H} \mathbf{y},
\end{align}
where \( \mathbf{A}^\mathrm{H} \) is called the Hermitian (conjugate transpose) of \( \mathbf{A} \).  
If \({A}\) is a scalar, its Hermitian conjugate simplifies to the complex conjugate, i.e., \( {A}^\mathrm{H} = \overline{{A}} \).\\
Let \( \mathbf{A} \in \mathcal{B}(\mathcal{H}) \) be an \( n \times n \) square operator. The trace of \( \mathbf{A} \), as defined in \cite[Definition 3.5.1]{huppert2010lineare}, is  
\begin{align}\label{TrDef}
    \operatorname{Tr}(\mathbf{A}) = \sum_{i=1}^{n} \mathrm{a}_{ii}, \quad\mathrm{where}\quad n\leq N.
\end{align}
If \( \mathbf{A} \) is not square, the trace is undefined.
\begin{theo}[Cyclic Property of the Trace {\cite[Definition~3.5.1, Satz~3.5.2]{huppert2010lineare}}]\label{trace_properties}
Let \( \mathbf{A}, \mathbf{B} \in \mathcal{B}(\mathcal{H}) \). If the product \( \mathbf{A} \mathbf{B} \) (or equivalently \( \mathbf{B} \mathbf{A} \)) is a square operator, then the trace satisfies the cyclic property:
\begin{equation}
    \operatorname{Tr}(\mathbf{A} \mathbf{B}) = \operatorname{Tr}(\mathbf{B} \mathbf{A}).
\end{equation}
\end{theo}

\textit{Proof.} Follows directly from~\cite[Theorem~3.5.2]{huppert2010lineare}. \hfill\(\blacksquare\)

\subsection{Semidefinite Relaxation Optimality }\label{Appen_SRO}
In this section, we examine the separable \ac{SDP} problem with different constraints. The problem can be formulated in the following general form~\cite[Section~III]{huang}:
\begin{subequations}\label{P1}
    \begin{align}
        \mathrm{(P1)} \quad & \max_{\mathbf{W}_1, \dots, \mathbf{W}_L} \quad \sum_{l=1}^{L} \operatorname{Tr}(\mathbf{C}_l \mathbf{W}_l),  \\
        & \text{s.t.} \quad \sum_{l=1}^{L} \operatorname{Tr}(\mathbf{A}_{ml} \mathbf{W}_l) \trianglelefteq_m b_m, \quad m = 1, \dots, M, \\
        & \phantom{\text{s.t.} \quad} \mathbf{W}_l \succeq 0, \quad l = 1, \dots, L, 
    \end{align}
\end{subequations}
where \( \mathbf{C}_l, \mathbf{A}_{ml} \in \mathcal{B}(\mathcal{H}) \) are positive semidefinite for all \( l \) and \( m \). The parameters \( b_m \in \mathbb{R} \), and the relation \( \trianglelefteq_m \in \{\geq, =, \leq\} \) indicates the type of constraint associated with index \( m \). The decision variables \( \mathbf{W}_l \in \mathcal{B}(\mathcal{H}) \) are Hermitian operators for all \( l \). This problem exhibits a well-known rank-reduction property, as stated in the following lemma:
\begin{lemma}[{\cite[Lemma~3.1]{huang}}]\label{lamm3.1}
Suppose that the separable \ac{SDP} problem~(P1) in~\eqref{P1} and its dual are both solvable. Then, there always exists an optimal solution 
\begin{align}
    (\mathbf{W}_1^\star, \dots, \mathbf{W}_L^\star) \in \left( \mathcal{B}(\mathcal{H}) \right)^L \notag
\end{align}
such that the following finite-rank constraint holds:
\begin{equation}
    \sum_{l=1}^{L} \operatorname{rank}^2(\mathbf{W}_l^\star) \leq M < \infty.
    \label{rankinq}
\end{equation}
\end{lemma}
\textit{Proof}: See \cite[Appendix A]{huang}. \hfill \(\blacksquare\)\\
Based on Lemma~\ref{lamm3.1}, we can constrain the rank of \( \mathbf{W} \) to a finite integer \( M \), providing a theoretical justification for limiting the rank of the beamforming operator. This constraint effectively reduces the solution search space.

\begin{enumerate}
   \item \textbf{Case \( \operatorname{rank}(\mathbf{W}^\star) = 1 \):} 
The problem becomes convex and can be efficiently solved via CVXPY~\cite{cvx1}, achieving global optimality in polynomial time, following convex optimization theory in Hilbert spaces $\mathcal{H}$~\cite{cvx2}.

\item \textbf{Case \( \operatorname{rank}(\mathbf{W}^\star) > 1 \):} 
Heuristic methods may extract a feasible \( \mathbf{w} \), but without global optimality; see~\cite{luo} for \ac{SDR} recovery techniques.

\end{enumerate}

\end{document}